\documentstyle[preprint,aps]{revtex}
\begin{document}
\preprint{DRAFT }
\title{Noncommutative Quantum Mechanics: The Two-Dimensional Central Field}
\author{J. Gamboa$^1$\thanks{E-mail: jgamboa@lauca.usach.cl}, M. Loewe
$^2$\thanks{E-mail: mloewe@fis.puc.cl}, F. M\'endez$^1$\thanks{E-mail: fmendez@lauca.usach.cl} and J.
C. Rojas$^3$
\thanks{E-mail: rojas@sonia.ecm.ub.es}}
\address{$^1$Departamento de F\'{\i}sica, Universidad de Santiago de Chile,
Casilla 307, Santiago 2, Chile \\
Facultad de F\'{\i}sica, Pontificia Universidad Cat\'olica de Chile, Casilla 306,
Santiago 22, Chile \\
$^3$Departament ECM, Facultat de Fisica, Universitat de Barcelona and Institut
D'Altes Energies,
\\
Diagonal 647, E-08028, Barcelona, Spain }
\maketitle
\begin{abstract}
Quantum mechanics in a noncommutative plane is considered. For a general two
dimensional central field,  we find that the theory can be perturbatively solved for
large values of  the noncommutative parameter ($\theta$) and explicit expressions for the
eigenstates and eigenvalues are given. The Green function  is explicitly obtained and
we show that it can be expressed as  an infinite series.  For  polynomial type
potentials, we found a smooth limit for  small values of $\theta$ and for non-polynomial ones this limit is
necessarily abrupt. The Landau
problem, as a limit case of a noncommutative system, is also  considered.
\end{abstract}
\narrowtext

\section{Introduction}

Recent results in string theory suggest that  spacetime could be noncommutative
\cite{string}. If this claim is correct, then important new implications in
our conception of  space and time could take  place\cite{varios}. For instance,
spatial noncommutativity implies new Heisenberg-type relations, namely
\begin{equation}
\Delta x \Delta y \sim \theta, \label{1}
\end{equation}
where $\theta$ is a measure of the noncommutative effects and plays an
analogous role of $\hbar$ in standard quantum mechanics\cite{hei}. The space-time
noncommutativity, however, violates causality at the quantum field theory level although
space-time noncommutativity could be consistent in string theory\cite{susk}.

The goal of this paper is to dicuss two dimensional  quantum mechanics in a
noncommutative space. More precisely, we will consider the central field case which,
remarkably,  can be solved {\it almost} completely and the solution shows
quite explicitly the difficulties and virtues found in noncommutative quantum field
theory. Thus, in this sense, our model could be considered as a framework
where one could explicitly check some properties that appear  in quantum field theory.

The paper is organized as follows: in section 2, we review the model proposed in
\cite{glr} and we discuss their properties and physical interpretation. In section 3 we
write a general expression for the Green function. Section 4 is devoted to  the
calculation of the partition function and the statistical mechanics of simple systems,
sketching  a perturbative procedure  for more general cases. Finally, in section 5  the
conclusion and an outlook is presented.

\section{Quantum mechanics in a noncommutative space}

Noncommutative quantum mechanics is a theory defined on a manifold where
the product of functions  is the Moyal one. If
$A({\bf x})$ and $B({\bf x})$ are two functions, then the Moyal product is defined
as
\begin{equation}
{\bf A}\star {\bf B}({\bf x}) = e^{\frac{i}{2}\theta^{ij}\partial^{(1)}_i \partial^{(2)}_j}
{\bf A}({\bf x}_1) {\bf
B}({\bf x}_2) \vert_{{\bf x}_1={\bf x}_2={\bf x}}. \label{2}
\end{equation}

This formula implies  that the Schr\"odinger equation
\begin{equation}
i \frac{\partial \Psi ({\bf x},t)}{\partial t} = \biggl[ \frac{{\bf p}^2}{2 m} + V ({\bf x})
\biggl] \Psi ({\bf x},t),
\label{3}
\end{equation}
in the noncommutative space is the same one but with the potential shifted as
$V({\bf x} -  \frac{{\tilde {\bf p}}}{2})$, where $\tilde p _{i}=\theta^{ij} p_{j}$,
$\theta _{ij} = \theta \epsilon _{ij}$ and  $\epsilon_{ij}$ is the antisymmetric tensor in
two dimensions. Although this result appeared in connection with string theory
\cite{sussk}  there is also an older version  known as Bopp's shift
\cite{bopp}. This last fact implies that quantum mechanics in a noncommutative plane
is highly nontrivial because, as the shifted potential involves in principle arbitrary
powers of the momenta,  we will have an arbitrary large number of derivatives in the
Schr\"odinger equation.

Thus,  the question is  how to handle  noncommutative systems in a simple way. In
order to give an answer to this question, let us consider a central field in
two dimensions $V({\bf x}) = V (r^2)$. In the noncommutative space this potential is
equivalent to the replacement
\begin{eqnarray}
V({\bf x}) \rightarrow &{}& V ( \frac{\theta^2}{4} p_x^2 + x^2 + \frac{\theta^2} {4}
p_y^2 + y^2  - \theta L_z)
\nonumber
\\
&=& V ({\hat {\aleph}}), \label{5}
\end{eqnarray}
where the aleph operator (${\hat \aleph}$)  is defined as
\begin{equation}
{\hat \aleph} = {\hat H}_{HO} - \theta{\hat L}_z, \label{7}
\end{equation}
where $H_{HO}$ is the hamiltonian for a two dimensional harmonic oscillator with
mass $2/\theta^2$,
frequency $\omega =
\theta$ and $L_z$ is the z-component of the angular momentum defined as
$L_z = x p_y - y p_x$.

The eigenstates and the eigenvalues of the $\aleph$ operator can be calculated
noticing that in two dimensions the harmonic oscillator is  associated to the $SU(2)$
group whose generators are given by
\begin{eqnarray}
L_x &=& \frac{1}{2} ( a^{\dagger}_x a_x - a^{\dagger}_y a_y), \nonumber
\\
L_y &=& \frac{1}{2} ( a^{\dagger}_x a_y + a^{\dagger}_y a_x), \label{8}
\\
L_z &=& \frac{1}{2i} ( a^{\dagger}_x a_y - a^{\dagger}_y a_x), \nonumber
\end{eqnarray}
and, furthermore, $(\aleph, {\hat L}^2, J_z = \frac{1}{2}L_z)$ is  a complete set of
commuting observables. For the calculation of the eigenvalues is more convenient to
use
\begin{eqnarray}
a_\pm &=& \frac{1}{\sqrt{2}}(a_y \pm i a_x), \nonumber
\\
a^\dagger_\pm &=& \frac{1}{\sqrt{2}}(a^\dagger_y \pm i a^\dagger_x). \label{9}
\end{eqnarray}

In the basis  $\vert n_+,n_->$,  the operators $(\aleph, {\hat L}^2, J_z )$ are diagonal
and the quantum numbers take the values $n_{\pm} =0,1,2,3...$. By using this, one
get  that the eigenvalues of $\aleph$ are
\begin{equation}
\lambda_{n_+n_-} = \theta \,[\,2 n_- + 1].\label{eigen}
\end{equation}
Note  that the spectrum  of $\aleph$ does not  depend on the label $n_+$ and their
eigenvalues are infinitely degenerated.

Remarkably, one can compute exactly the eigenvalues associated to $V(\aleph)$.
Indeed, let $a_n$ be the eigenvalues associated to the operator ${\hat A}$ and
$\psi_n$ the corresponding eigenfunction.  Then for a small shift of the argument of an
arbitrary  function $f({\hat A})$ one find that
\begin{equation}
f({\hat A} + \epsilon) \psi_n = f(a_n + \epsilon)\psi_n. \label{10}
\end{equation}
By using (\ref{10}), the following identity is found
\begin{equation}
V(\aleph) \vert n_+,n_-> = V(\theta \,[2 n_- +1]) \vert n_+,n_->. \label{11}
\end{equation}

Once equation (\ref{11}) is obtained, we can compute the eigenvalues and
eigenfunctions for a general system described by the hamiltonian
\begin{equation}
{\hat H} = \frac{1}{2 M} {\bf p}^2 + V({\bf x}^2),\label{111}
\end{equation}
which can be rewritten as follows:
\begin{eqnarray}
{\hat H }&=& \frac{{\bf p}^2}{2M} + V(\hat{\aleph}), \label{land}  \nonumber
\\
&=&  \frac{2}{M \theta^2} \left(  \frac{\theta^2 }{4} {\bf p}^2  + {\bf r}^2 + V(\hat{\aleph})
\right) - \frac{2}{M \theta^2} {{\bf r}^2}   \nonumber
 \\
 &\equiv& H_0 - \frac{2}{M \theta^2} {\bf r}^2 . \label{ori}
 \end{eqnarray}

 Then, after  using  (\ref{11}), the eigenvalues $\Lambda_{n_+,n_-}$ of
 $H_0$ are
 \begin{equation}
 \Lambda_{n_+n_-} = \frac{2}{M \theta} \left [  n_+ + n_- +1 \right]
 + V \left[ \,\theta \, (2n_-+1)\right],
\end{equation}
while the eigenvalues of ${\hat H}$ are
\begin{eqnarray}
E_{n_+n_-}& = & <n_+n_-\vert \hat{H}_0 \vert n_+n_-> - \frac{2}{M \theta^2}<n_+n_-
\vert {\bf r}^2 \vert n_+n_->   \nonumber
\\
&=& \frac{2}{M \theta} [ n_++n_- +1 ] + V[\, \theta \,(2n_- +1)] - \frac{2}{M \theta^2}  <n_
+
n_-\mid  { {\bf r}^2}  \mid n_+n_-> .
\label{formula}
\end{eqnarray}

In this expression, the last term of the {\it R.H.S.} can be computed by using
perturbation theory for large values of $\theta$ and, as a consequence, the
eigenfunctions that appear  in the matrix element
$< n_+n_-  \mid  { {\bf r}^2}  \mid n_+n_->$, are those of the  two dimensional
harmonic oscillator, {\it i.e.}
\begin{equation}
\vert n_+n_-> = \frac{{a_+^\dagger}^{n_+} {a_-^\dagger}^{n_-}}{\sqrt{(n_+)! (n_-)!}}
\vert 0,0>. \label{os}
\end{equation}

Thus, in this approximation the eigenvalues that correspond to the diagonal part of the
hamiltonian are
\begin{equation}
\tilde{E}_{n_+n_-} = \frac{1}{M\theta} (n_++n_--1) + V(\theta[2n_-+1]). \label{16}
\end{equation}

Although the higher order corrections to the eigenfunctions and eigenvalues are
straightforward calculable  the perturbative series cannot be expressed in a closed
way.

One should note that quantum mechanics in a noncommutative space  has new
consequences. In particular,  the usual Schr\"odinger equation in the commutative
case
\begin{equation}
(\triangle + k^2 ) \psi ({\bf x}) = V({\bf x}) \psi ({\bf x}), \label{4}
\end{equation}
has an  integral version
\begin{equation}
\psi (x) = \varphi (x) + \int dx^{'} G[x,x^{'}] V(x')\psi (x^{'}) , \label{green}
\end{equation}
which cannot be realized in the noncommutative space, since the
shift $x \rightarrow x + \frac{1}{2} \theta \nabla$ in the potential involves unknown
powers  of  the derivatives. These higher powers could spoil the unitarity (conservation
of probability) of quantum mechanics in a noncommutative space. However, this is
circumvented by the properties satisfied by  the Moyal product. These  imply that the
equation
\begin{equation}
\nabla \cdot {\bf J} = -\frac{\partial \rho}{\partial t}, \label{conti}
\end{equation}
with $\rho = \psi^\ast \psi$  and $\psi^* \nabla \psi - \psi \nabla \psi^+$ remains valid
\cite{rivelles}  and, therefore, the physical meaning of the wave function in a
noncommutative space does not change.

\section{Green Functions in Noncommutative Spaces}

In this section we will discuss the calculation of the Green function
$G[{\bf x}, {\bf x}^{'}]$ for the previous model in a $1/\theta$ power expansion. For
$\theta>>1$, we can   compute $G[{\bf x}, {\bf x}^{'}]$ to any order in $1/\theta$ since
we need  the matrix element $<n_+n_-\vert {\bf r}^2\vert n_+^{'} n_-^{'}>$.

For $\theta >>1$, as we discussed above,  the term ${\bf r}^2$ is a  non-diagonal
perturbation and the 0-order wave function are those of the two dimensional
harmonic oscillator, {\it i.e.} the  basis $\vert n_+n_->$. Thus, assuming this condition,
the Green function is given by
\begin{eqnarray}
G[{\bf x}, {\bf x}^{'}] &=& <{\bf x}\vert e^{-i {\hat H} (t-t^{'})}\vert {\bf x}^{'}>
\nonumber
\\
&=& \sum_{n_+n_-,n_+^{'}n_-^{'}} <{\bf x}\vert n_+n_-><n_+n_-\vert e^{-i {\hat H}
(t-t^{'})} \vert n_+^{'}n_-^{'}><n_+^{'}n_-^{'}\vert {\bf x}^{'}>,
\nonumber
\\
&=& \sum_{n_+n_-,n_+^{'}n_-^{'}} \psi^\ast_{n_+n_-} ({\bf x})\,\psi_{n_+^{'}n_-^{'}} ({\bf
x}^{'}) <n_+n_-\vert e^{-i {\hat H} (t-t^{'})} \vert n_+^{'}n_-^{'}>, \label{green1}
\end{eqnarray}
where the wave function $\psi_{n_+n_-} ({\bf x})$ are the usual Hermite functions.

As ${\bf r}^2$ does not commute with $\aleph$ and $L_z$, the matrix element
$<n_+n_-\vert e^{-i {\hat H} (t-t^{'})} \vert n_+^{'}n_-^{'}>$ must
be calculated by expanding the evolution operator $e^{-i {\hat H} (t-t^{'})}$, {\it i.e.}
\begin{equation}
<{\bf x}\vert e^{-i {\hat H} (t-t^{'})}\vert {\bf x}^{'}>= \sum_{k=0}^{\infty} \frac{(-i)^k}{k!}
\,\, <x\vert  [H (t-t^{'})]^k \vert x^{'}>. \label{exp}
\end{equation}

If we use recursively the completeness relation for $\vert n_+n_->$, we obtain
\begin{eqnarray}
<{\bf x}\vert &e&^{-i {\hat H} (t-t^{'})}\vert {\bf x}^{'}>= \sum_{k=0}^{\infty}\, \sum_{n_+^{(
1)}n_-^{(1)}...n_+^{(k)}n_-^{(k)}} \frac{(-i)^k}{k!} \times \nonumber
\\
&\times& <n_+n_-\vert {\hat H}\vert n_+^{(1)}n_-^{(1)}> <n_+^{(1)}n_-^{(1)}\vert {\hat H}
\vert n_+^{(2)}n_-^{(2)}>...<n_+^{(n-1)}n_-^{(k-1)}\vert {\hat H}\vert n_+^{'}n_-^{'}>
{(t-t^{'})}^k. \label{exp2}
\end{eqnarray}

Using the hamiltonian (\ref{ori}) and the properties of the operators $\aleph$ and
$L_z$, then each bracket in (\ref{exp2})  becomes
\begin{eqnarray}
<n_+n_-\vert {\hat H} \vert n_+^{'}n_-^{'}> &=& \tilde{E}_{n_+n_-} \delta_{n_+,n_+^{'}}
\delta_{n_-,n_-^{'}} - \frac{1}{M \theta} [ \sqrt{(n_+^{'}+1)(n_-^{'}+1)} \delta_{ n_+,n_+^{'}
+ 1} \delta_{n_-,n_-^{'}+1} \nonumber
\\
&+& \sqrt{n_+^{'}n_-^{'}}\delta_{n_+,n_+^{'}-1 }\delta_{n_-,n_-^{'}-1}],
\label{correc}
\end{eqnarray}
where $\tilde{E}_{n_+n_-}$ is defined in (\ref{16}).

In order to simplify  the  calculations,  it is better to rewrite (\ref{correc}) in terms of the
variables $j,m$ defined as $2j=n_-+n_+$ and $m=n_+-n_-$, respectively,  which  take
the values $j=0,\frac{1}{2},1,\frac{3}{2}...$ and $m=-j,-j+1...,j-1,j$.

The result  is\footnote{This notation must be taken with care since here we are not
working with the usual basis in  polar coordinates.}
\begin{equation}
<n_+n_-\vert {\hat H} \vert n_+^{'}n_-^{'}> = \tilde{E}_{jm} \delta_{jj^{'}} \delta_{mm^{'}} -
\frac{1}{M\theta} [ \sqrt{j^2-m^2} \delta_{j-1 j^{'}} + \sqrt{{(j+1)}^2 -m^2}\delta_{j+1j^{'}}] ,
\label{correc2}
\end{equation}
and, as a consequence, the Green function becomes
\begin{eqnarray}
G[x,x^{'}] &=&\sum_{j,m}\biggl[ \psi _{jm}^{*}(x) \psi_{jm}(x^{\prime})S_{jm} +
\nonumber
\\
&& \left( -\frac{1}{M\theta }\right) \left( \psi_{j-1,m}^{*}(x^{\prime })\psi _{jm}(x)S_{j-1,m}
+ \psi _{j+1,m}^{*} (x^{\prime}) \psi _{jm}(x)S_{j+1,m}\right) +\nonumber
\\
& &{\left( -\frac{1}{M\theta }\right)}^{2}
\left( \psi_{j-2,m}^{*}(x^{\prime })\psi _{jm}(x)S_{j-2,m}+\psi _{j+2,m}^{*}(x^{\prime})
\psi _{jm}(x)S_{j+2,m}\right) +...\biggr], \label{gre1}
\end{eqnarray}
where the $S_{jm}$ are given by (with $\tau =-it$)
\begin{eqnarray}
S_{jm} &=&e^{-\tau \tilde{E}_{jm}}+\left( \frac{\tau }{M\theta }\right)
^{2}(j^{2}-m^{2})+\left( \frac{\tau }{M\theta }\right) ^{3}\left[
.....\right] , \label{s1}\\
S_{j-1,m} &=&\sqrt{j^{2}-m^{2}}\left[ \tau +\tau ^{2}\left(
\tilde{E}_{j-1}+\tilde{E}_{j}\right) +\tau ^{3}\left(
\tilde{E}_{j-1}^{2}+\tilde{E}_{j-1}\tilde{E}_{j}+\tilde{E}_{j}^{2}\right) +....\right] , \\
S_{j+1,m} &=&\sqrt{(j+1)^{2}-m^{2}}\left[ \tau +\tau ^{2}\left(
\tilde{E}_{j+1}+\tilde{E}_{j}\right) +\tau ^{3}\left(
\tilde{E}_{j+1}^{2}+\tilde{E}_{j+1}\tilde{E}_{j}+\tilde{E}_{j}^{2}\right) +....\right] , \\
S_{j-2,m} &=&\sqrt{j^{2}-m^{2}}\sqrt{(j-1)^{2}-m^{2}}\left[ \tau
^{2}+\tau ^{3}\left( \tilde{E}_{j-2}+\tilde{E}_{j-1}+\tilde{E}_{j}\right) +....\right] , \\
S_{j+2,m} &=&\sqrt{(j+2)^{2}-m^{2}}\sqrt{(j+1)^{2}-m^{2}}\left[ \tau
^{2}+\tau ^{3}\left( \tilde{E}_{j+2}+\tilde{E}_{j+1}+\tilde{E}_{j}\right) +....\right].
\end{eqnarray}

This is the full Green function for the motion of a particle in a general central field
if $\theta >>1$. Thus, in a perturbative sense, our model is exactly solvable although
we cannot find a closed form  for the perturbative series.

Notice  that  the potential contains dynamical information induced by the presence of
the momentum due to the shift $x \rightarrow x - \tilde{{\bf p}}/2$,  allowing then to
consider the kinetic term ${\bf p^2}/2M$ as a perturbation.

In order to do that,  one must evaluate the matrix element
$<n_+n_- \mid{\bf p}^2\mid n_+n_->$ as  was done previously for
${\bf r}^2$. The result is
\begin{eqnarray}
<n_+n_- \mid{\bf p}^2\mid n_+&n&_->= \frac{2}{\theta}(n_-+n_++1)\delta_{n_+,n'_+}
\delta_{n_-,n'_-} \nonumber
\\
&-& \frac{2}{\theta} [\sqrt{(n'_++1)(n'_-+1)}\delta_{n_+,n'_++1}\delta_{n_-,n'_-+1}
+\sqrt{n'_+n'_-}\delta_{n_+,n'_+-1}\delta_{n_-,n'_--1} ],
\label{pcua}
\end{eqnarray}
which gives the same expresion for the Green function as equation (\ref{gre1}).

It is interesting to notice,  at  this point,  that the mass $M$ and the noncommutative
parameter $\theta$ appear   always in the combination  $1/(M\theta)$, and as a
consequence,  our calculation is insensitive to the replacement
$M \leftrightarrow \theta$;
this sort of duality explains why the limits $M \rightarrow \infty$ and
$\theta \rightarrow \infty$ give  the same results at perturbative level.

\section{The Landau Problem as a Noncommutative System}

In this section we will explain how the Landau problem can be understood as a
noncommutative one \cite{GLMR}. Although this idea has been discussed recently
\cite{poly}, there is an older  version on this problem\cite{belli}. Following the results
given in section II,  if we choose  the potential $V(\aleph)$ in (\ref{land}) as
\begin{equation}
V(\aleph) =  \Omega \,\aleph. \label{oa}
\end{equation}
where $\Omega$ is an appropriate constant, then the hamiltonian becomes
\begin{equation}
H = (\frac{1}{2M} + \frac{ \Omega\, \theta^2}{4}) {\bf p}^2 + \Omega\, {\bf x}^2 - \Omega
\, \theta L_z. \label{land1}
\end{equation}

This hamiltonian is diagonal in the basis  $\vert n_+n_->$ and\footnote{This basis is
the same one  where $\aleph$ is diagonal. However, in the coordinates
representation,  they look  due to the definition of the operators
$a_{\pm},a_{\pm}^{\dag}$.  For details see the Appendix.}
\begin{equation}
H\vert n_+n_->=\bigg[\sqrt{\frac{2\Omega}{{\cal M}}} \bigg(n_++n_-+1\bigg) -\Omega
\theta(n_+-n_-) \bigg] \vert n_+n_->,
\label{hamtotal}
\end{equation}
where ${\cal M}$ is an effective mass defined as
$\frac{1}{{\cal M}} = \frac{1}{M} + \frac{\Omega \theta^2}{2}$.

The propagator can be computed as in the previous section, {\it but now for any value
of  $\theta$}. Indeed, the Green function becomes the matrix element
$<x\vert e^{-i[\frac{p^2}{2M}+ \Omega\,\aleph] T}\vert x^{'}>$ and the
explicit calculation gives
\begin{eqnarray}
G[x,x^{'}; T] &=& <x\vert e^{-i[\frac{p^2}{2M}+ \Omega\,\aleph] T}\vert x^{'}>, \nonumber
\\
&=& \sum_{n_+n_-} \psi_{n_+n_-}^* ({\bf x}) \psi_{n_+n_-} ({\bf x}^{'}) \,e^{- i [\sqrt{
\frac{2\Omega}{{\cal M}}} (n_++n_-+1) -\Omega \theta(n_+-n_-) ] T }, \label{para1}
\end{eqnarray}
where $T=t-t'$ and $\psi_{n_+n_-} ({\bf x})$ are the eigenfunctions of the harmonic
oscillator in cartesian coordinates.

From (\ref{para1}) one can compute the associated partition function
\begin{equation}
Z=\mbox{Tr} G(x,x')
\label{partition}
\end{equation}
once the euclidean rotation $iT \rightarrow \beta$  has been performed. Using  this,
one can see that the usual formula in the commutative space
\begin{equation}
Z=\sum_{n_+,n_- =0} e^{-\beta E_{n_+n_-}},
\label{usual}
\end{equation}
hold  in the noncommutative case. It is interesting to note that for a general
$V(\aleph)$,
equation (\ref{gre1}) yields a partition function like  (\ref{usual}) only in the large
$\theta$ limit; otherwise, new non exponential contributions  should be added.

Finally, the partition function is
\begin{eqnarray}
Z &=& \sum_{n_+,n_-=0} \exp\bigg(-\beta\bigg[ \bigg( \sqrt{\frac{2\Omega}{{\cal M}}} -
\Omega \theta \bigg) n_+ -\bigg( \sqrt{\frac{2\Omega}{{\cal M}}} +\Omega \theta \bigg)
n_- + \sqrt{\frac{2\Omega}{{\cal M}}}\bigg] \bigg), \label{pa0} \\
&=& \frac{1}{4 \sinh \bigg[\beta \bigg(\sqrt{\frac{2\Omega}{{\cal M}}} -\Omega \theta
\bigg)\bigg] \sinh \bigg[\beta\bigg( \sqrt{\frac{2\Omega}{{\cal M}}} +\Omega
\theta\bigg) \bigg] }
\label{pa}
\end{eqnarray}
which is the usual partition function of the two dimensional harmonic oscillator.

Now we can map (\ref{land1}) into the Landau problem. If the magnetic field
${\bf H} = H_0 {\bf e}_3$, then the hamiltonian  in the symmetric gauge for a
particle with mass $\mu$
\begin{equation}
{\hat H}_{\mbox{Landau}} = \frac{1}{2\mu} {\bf p}^2 + \frac{e^2H^2_0}{8\mu^2} {\bf
x}^2  - \frac{eH_0}{2 \mu}
L_z, \label{a1}
\end{equation}
 just coincides with (\ref{land1}) if we identify
\begin{eqnarray}
\frac{1}{2\mu}&=& \frac{1}{2M} + \frac{\Omega\, \theta^2}{4},\,\,\,\,\,\,\,\,\,\, \frac{e^2H^2_
0}{8 \mu}=\Omega, \label{para}
\\
\Omega \,\theta &=& \frac{eH_0}{2 \mu}. \label{theta}
\end{eqnarray}

These equations are consistent  if and only if $M=\infty$, {\it i.e.} the equivalence
between noncommutative quantum mechanics and the Landau problem is exact for
the lowest Landau level.

Thus
\begin{eqnarray}
\Omega &=& \frac{e^2 H_0}{8 \mu}, \nonumber
\\
\theta&=&\frac{4}{e H_0}.
\label{relaciones}
\end{eqnarray}

A  simple  dimensional analysis shows that $\Omega$ has [energy/length$^2$]
dimensions and,  therefore is  enough to change $\theta$ by
$\tilde{\theta}=\hbar \theta$ in order to have a noncommutative parameter with the
appropriated dimensions [length]$^2$.

Considering magnetic fields in the region of quantum Hall effect ($\sim$ 12 T) one
can find a bound for the noncommutative parameter {\it i.e.}
\begin{equation}
\bar{\theta} = 0.22\times 10^{-11} \mbox{cm}^2.
\end{equation}
For this value of $\tilde{\theta}$ one cannot distinguish between noncommutative
quantum mechanics and the Landau problem.

The partition function in the  limit $M \rightarrow \infty$ can be computed by noticing
that  the argument of the first  $sinh$ in the denominator vanishes and therefore the
partition function diverges. This remind us  that in (\ref{pa0}) we had an infinite sum
over $n_+$  which must be regularized.

The partition function for the Landau problem becomes
\begin{equation}
Z=\frac{{\cal N}}{2\sinh(\frac{\beta e H_0}{2\mu})},
\label{oscillator}
\end{equation}
where  ${\cal N}$ is an infinite constant that is irrelevant for the thermodynamics analysis, although can
computed by using $\zeta$-function regularization  giving ${\cal N} =1/2$.

The magnetization and the magnetic susceptibility yield to the usual expressions \cite{landau}, {\it i.e.}
\begin{eqnarray}
M&=& - \frac{4 e}{ \mu} \coth(\frac{\beta e H_0}{2\mu}) \nonumber
\\
\chi &=&-\frac{16 e^2 \beta}{\mu^2}\mbox{cosech}\bigg[\frac{4 H_0 e \beta}{\mu}
\bigg]
\label{diamgnetos}
\end{eqnarray}
and,  therefore,  the system is diamagnetic.

\section{Conclusions}
In this paper, noncommutative quantum mechanics for a  two dimensional central field
was considered. The main point is that  we have established a mapping between any
noncommutative quantum mechanical system in two dimensions with a commutative
version. In the commutative space the potential is shifted by
$V({\bf x} - \frac{1}{2} {\tilde {\bf p}})$, where $V({\bf x})$ is the potential defined on the
noncommutative space. This last fact is a consequence of the Moyal product or, in other
words, the Moyal product  provides an explicit realization of the Seiberg-Witten map\cite{string}.

In the central field case, one can compute the spectrum of the hamiltonian from
(\ref{16}). Indeed, if $V({\bf x})$  has the form  $ x^m$, then if $m>0$
\begin{equation}
\tilde{E}_{n_+n_-}=\frac{2}{M\theta} (n_++n_-+1) -\frac{2}{M\theta^2}<n_+n_-\mid {\bf r}^2
\mid n_+n_->+ [\theta (2n_- +1)]^{m/2}.
\label{polinomios}
\end{equation}

If $\theta >>1$ the second term is computed by using perturbation theory and, the
diagonal part of (\ref{polinomios}) becomes
\begin{equation}
\tilde{E}_{n_+n_-}=\frac{1}{M\theta} (n_++n_-+1) +  [\theta (2n_- +1)]^{m/2},
\label{polidos}
\end{equation}
and, therefore, the dominant part of the spectrum becomes $[\theta (2n_- +1)]^{m/2}$, {\it i.e.} is
infinitely degenerated in $n_+$.

If  $\theta << 1$, only  the first two terms in (\ref{polinomios}) are the dominant contributions.

If $m<0$ but $\vert m\vert >2$, then the main contributions to $\tilde{E}_{n_+n_-}$ comes from the terms
proportional to $1/M$ . When $\vert m\vert <2$, the dominant contributions are due to
$[\theta (2n_- +1)]^{m/2}$.

If the potential is non-polynomial, then the limit $\theta<<1$ could not exist as, {\it e.g.} the Coulomb
potential.
\acknowledgments
We would like to thank Prof. V.O. Rivelles by comments on the manuscript. This work
was partially supported by FONDECYT-Chile under grant numbers 1010596, 1010976 and 3000005.

\appendix
\section{}
In this section we set the definition of the operators $a_{\pm},a^\dag_{\pm}$,  in order
to construct the basis $\mid n_+n_->$.

The operator $\aleph$ can be diagonalized as follow; defining
\begin{eqnarray}
\hat{a}_{x}&=&\frac{1}{\sqrt{\theta}}(\hat{x} +\frac{i}{2\theta} \hat{p}_x),
\\
\hat{a }^\dag_{x}&=&\frac{1}{\sqrt{\theta}}(\hat{x} -\frac{i}{2\theta} \hat{p}_x).
\label{aes}
\end{eqnarray}
This operators satisfy  the usual commutation relations
$[\hat{a}_i,\hat{a}_j^{\dag}]=\delta_{ij}$ and $[\hat{a}_i,\hat{a}_j]=0$.

Although the aleph operator can be written in terms of these operators, it assumes the
diagonal form only in the basis $\mid n_+,n_->$ generated by
\begin{eqnarray}
\hat{a}_{+}&=&\frac{1}{\sqrt{2}}(\hat{a}_y + i \hat{a}_x),
\\
\hat{a}_{-}&=&\frac{1}{\sqrt{2}}(\hat{a}_y - i \hat{a}_x),
\label{amasmenos}
\end{eqnarray}
due to the presence of $L_z$. The aleph can be written as
\begin{equation}
\aleph=\theta(2\hat{a}_-^{\dag} \hat{a}_- +1).
\label{diago}
\end{equation}

For the hamiltonian $\hat{H}$ given in the equation (\ref{land1}) the previous
definitions  (\ref{aes}) must be modified in order to have a diagonal form.

If we define
\begin{eqnarray}
\hat{a}_{x}&=&\frac{(2\Omega\mu)^{1/4}}{\sqrt{2}}(\hat{x} + \frac{i}{\sqrt{2}(2\Omega
\mu)^{1/4} } \hat{p}_x),
\\
\hat{a }^\dag_{x}&=&,\frac{(2\Omega\mu)^{1/4}}{\sqrt{2}}(\hat{x} - \frac{i}{\sqrt{2}(2
\Omega\mu)^{1/4} } \hat{p}_x)
\label{aes2}
\end{eqnarray}
and the corresponding $a_{\pm}$ defined in (\ref{amasmenos}), the hamiltonian
(\ref{land1}) turn out to be
\begin{equation}
\hat{H}=\sqrt{\frac{2\Omega}{\mu}}[\hat{a}_+^{\dag} \hat{a}_+ +\hat{a}_-^{\dag}
\hat{a}_- +1 ] - \Omega \theta [\hat{a}_-^{\dag}\hat{a}_- - \hat{a}_+^{\dag} \hat{a}_+].
\label{hamdiag}
\end{equation}

\end{document}